\documentclass[twocolumn]{aastex631}

\usepackage{booktabs}
\usepackage{minted}
\usepackage{threeparttable}
\usepackage{placeins}

\begin{document}

\title{An SMA Molecular Inventory of the Edge-on Protoplanetary Disk Gomez's Hamburger}

\author[0009-0002-8256-272X]{Erin M. Cusson} 
\affiliation{Department of Physics, Massachusetts Institute of Technology, Cambridge, MA 02139, USA}
\affiliation{Department of Earth, Atmospheric, and Planetary Sciences, Massachusetts Institute of Technology, Cambridge, MA 02139, USA}

\author[0000-0002-7212-2416]{Lisa W\"{o}lfer} 
\affiliation{Department of Earth, Atmospheric, and Planetary Sciences, Massachusetts Institute of Technology, Cambridge, MA 02139, USA}

\author[0000-0003-1534-5186]{Richard Teague}
\affiliation{Department of Earth, Atmospheric, and Planetary Sciences, Massachusetts Institute of Technology, Cambridge, MA 02139, USA}

\author[0000-0002-4248-5443]{Joshua B. Lovell}
\affiliation{Center for Astrophysics $\vert $ Harvard \& Smithsonian, 60 Garden Street, Cambridge, MA 02138, USA}

\author[0000-0001-8642-1786]{Chunhua Qi}
\affiliation{Institute for Astrophysical Research, Boston University, 725 Commonwealth Avenue, Boston, MA 02215, USA}

\author[0000-0003-2253-2270]{Sean M. Andrews}
\affiliation{Center for Astrophysics $\vert $ Harvard \& Smithsonian, 60 Garden Street, Cambridge, MA 02138, USA}

\author[0000-0002-9593-7618]{Thomas J. Haworth}
\affiliation{Astronomy Unit, School of Physics and Astronomy, Queen Mary University of London, London E1 4NS, UK}

\author[0000-0003-1008-1142]{John D. Ilee}
\affiliation{School of Physics and Astronomy, University of Leeds, Leeds, LS2 9JT, UK}

\author[0000-0001-6684-6269]{Marija R. Jankovic}
\affiliation{Institute of Physics Belgrade, University of Belgrade, Pregrevica 118, 11080 Belgrade, Serbia}

\author[0000-0003-1413-1776]{Charles J. Law}
\altaffiliation{NASA Hubble Fellowship Program Sagan Fellow}
\affiliation{Department of Astronomy, University of Virginia, Charlottesville, VA 22904, USA}

\author[0000-0003-1837-3772]{Romane le Gal}
\affiliation{Universit\'{e} Grenoble Alpes, CNRS, IPAG, F-38000 Grenoble, France}
\affiliation{Institut de Radioastronomie Millim\'{e}trique, 300 rue de la Piscine, Domaine Universitaire, 38406 Saint-Martin d’H\`{e}res, France}

\author[0000-0001-8798-1347]{Karin I. \"{O}berg}
\affiliation{Center for Astrophysics $\vert $ Harvard \& Smithsonian, 60 Garden Street, Cambridge, MA 02138, USA}

\author[0000-0003-1526-7587]{David Wilner}
\affiliation{Center for Astrophysics $\vert $ Harvard \& Smithsonian, 60 Garden Street, Cambridge, MA 02138, USA}

\begin{abstract}
Gomez's Hamburger (IRAS 18059-3211, GoHam) is a massive, edge-on protoplanetary disk that is potentially gravitationally unstable and hosts an overdensity that may be the site of a forming giant planet, making it a particularly interesting source for the study of planet formation in the direct collapse scenario. In this study, we present a molecular inventory of GoHam's disk combining several Submillimeter Array observations for a wideband survey at an angular resolution on the order of $\sim1\arcsec$. We detect 11 different molecules, including 15 individual lines, and measure their disk-integrated fluxes. We also infer column densities for several species over a range of fixed excitation temperatures. We find that the molecular inventory of GoHam and the inferred column densities for select molecules are broadly consistent with the general population of large protoplanetary disks. We explore the putative gravitational instability (GI) in GoHam's disk via possible enhancements in the gas-phase H$_2$CO abundance, but find no definitive evidence of GI. The results of this study can guide future, higher-resolution studies of GoHam, as well as efforts to characterize the giant protoplanet candidate GoHam b.
\end{abstract}

\keywords{Protoplanetary Disks; Chemical Abundances; Planet Formation} 

\section{Introduction}
\label{sec:intro}

It is generally accepted that giant planet formation occurs primarily through two main scenarios: core accretion  \citep[CA;][]{1972epcf.book.....S,1973ApJ...183.1051G,1985prpl.conf.1100H,1980PThPh..64..544M,1996Icar..124...62P} and gravitational instability \citep[GI;][]{1951PNAS...37....1K,1997Sci...276.1836B}. CA involves the accretion of planetesimals \citep{1978Icar...35....1G, 1996Icar..124...62P, 1998Icar..131..171K, 2011epsc.conf..242K} and pebbles \citep{2010A&A...520A..43O, 2014A&A...572A.107L, 2017AREPS..45..359J, 2023ASPC..534..717D}, creating a core that efficiently accretes gas as it becomes more massive. This is expected to be the more common of the two scenarios \citep{2007ApJ...662.1282M}, and can account for the formation of terrestrial planets and lower-mass giant planets (mass $<4M_{\mathrm{Jup}}$) \citep{2017A&A...603A..30S}. In higher-mass regimes, it becomes more difficult to explain the existence of giant planets, as the time required for them to form via CA may exceed the lifetime of the disk, especially around higher-mass stars \citep{2001ApJ...553L.153H}. Thus, an alternate planet formation pathway, i.e., GI, is required to explain more massive giant planets. In the GI scenario, instability within disks can cause fragmentation into self-gravitating clumps that collapse directly into giant gaseous protoplanets. This pathway is capable of producing giant planets on a much shorter timescale (on the order of $\approx10^3$ yr for $M_{\mathrm{disk}}/M_{\mathrm{star}}\approx0.1$) \citep{2000ApJ...536L.101B}. HR 8799 is an example of a system for which CA alone struggles to explain the observed planets, whereas GI has been shown to be a plausible pathway for their formation \citep{2010MNRAS.406.2279M}.

The GI scenario is particularly relevant to direct imaging observations that favor the detection of massive planets at large orbital separations ($>10$ au) from their host star.
Opportunities to study the disk conditions associated with the GI scenario of planet formation are rare, as the disk must meet the conditions necessary to produce gravitational instabilities that lead to direct collapse \citep[i.e. a massive disk or a favorable disk-to-star mass ratio][]{2016ARA&A..54..271K, 2020MNRAS.492.5041C}. Gomez's Hamburger (IRAS 18059-3211; GoHam) may be one such disk. With a disk mass of around 0.2 $M_{\odot}$ and a central star mass of $\approx 2.5$ $M_{\odot}$, the GoHam system's disk-to-star mass ratio is around an order of magnitude above average \citep{2011ARA&A..49...67W}, and with $^{12}$CO emission extending out to 1500 au, it is among the largest and most massive known protoplanetary disks \citep{Wood_2008,2010A&A...523A..18D} (see \citet{2024ApJ...967L...3B, 2024ApJ...967L...2M,2006ApJ...645.1272P} for discussion of similar rare sources). \citet{2020MNRAS.492.5041C} find that disks around higher-mass stars (i.e. M$_*\geq2$ M$_{\odot}$) are more susceptible to fragmentation and GI, and show that as stellar mass increases, the critical disk-to-star mass ratio needed for the occurrence of fragmentation decreases. \citet{2015A&A...578L...8B} estimate a Toomre $Q$ parameter of $Q\lesssim2$ for GoHam, which is consistent with marginal gravitational instability \citep{1964ApJ...139.1217T}. Thus, GoHam may be within the GI regime and is an interesting target for the study of disk conditions present in this planet formation scenario. 

\begin{figure*}[ht]
  \centering
  \includegraphics[width=\textwidth]{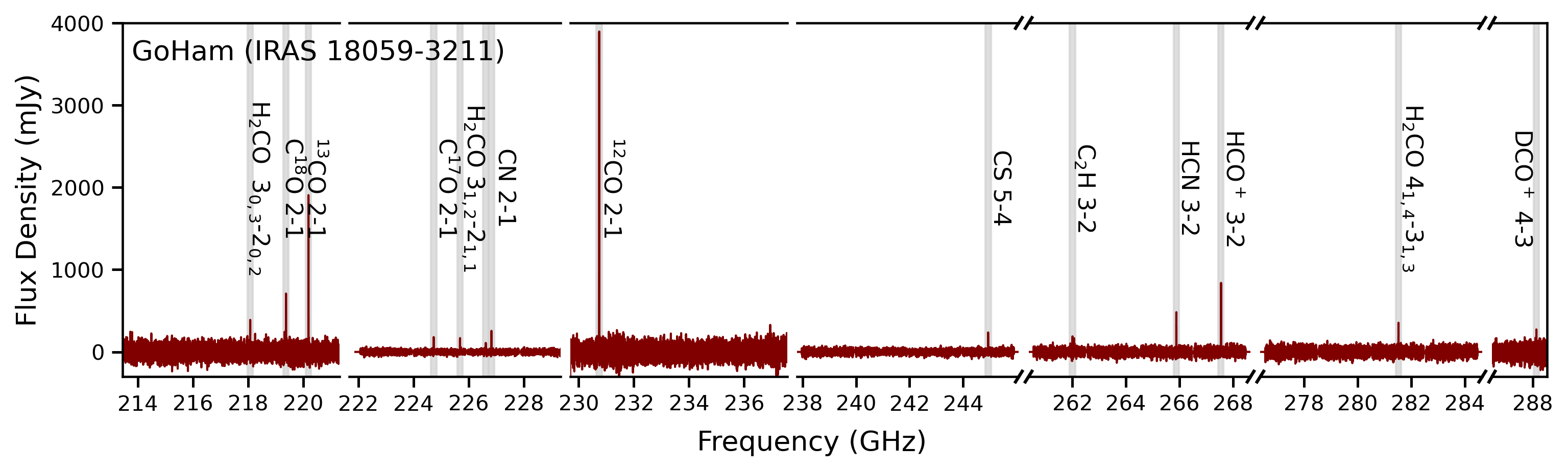}
  \caption{Summary of the frequency ranges observed and molecular lines (highlighted by vertical gray bars) detected within the SMA data. Note that the RMS is non-uniform across the spectrum as several different datasets were used.}
  \label{fig:figure1}
\end{figure*}

As we seek to refine our understanding of the planetary nurseries that give rise to planets through GI, it is also particularly useful to study sources which are viewed edge-on.
Highly-inclined viewing angles enable tracing of the physical and chemical structure of the disk in the vertical $(r,\,z)$ plane -- information that is more difficult to access in less-inclined sources. At an inclination of $i \approx 85^{\circ}$, GoHam is among these highly-inclined sources \citep{2008A&A...483..839B,2009A&A...500.1077B,Wood_2008,2010A&A...523A..18D, 2020MNRAS.495..451T}. 

The conditions present within GoHam may allow it to host particularly massive protoplanets at large orbital radii. \citet{2025ApJ...989....5T} showed correlations between the gas mass and the dust disk size which may suggest that more massive disks are capable of forming substructures farther from the central star. Indeed, GoHam contains a significant overdensity in its southern region at a radius of 350 au with a mass between 1 and 11 $M_{\mathrm{Jup}}$ \citep{2009A&A...500.1077B, 2015A&A...578L...8B, 2020MNRAS.495..451T}. This overdensity is suggested to be the site of a forming giant planet, dubbed GoHam b, that is theorized to have formed through GI \citep{2008A&A...483..839B, 2009A&A...500.1077B,2015A&A...578L...8B}.

Characterizing disk chemistry by identifying which molecules are present and abundant is 
crucial for our understanding of planet formation. In that regard, understanding how disk mass may affect disk chemistry and relative abundances of molecules is an important step. 
A rare disk such as GoHam, being massive, large, and potentially self-gravitating, may have different gas-phase molecular abundances than the general population of protoplanetary disks due to the effects of shock heating \citep{2011MNRAS.417.2950I, 2015MNRAS.453.1147E}. 
In this work, we present a molecular inventory of GoHam as observed with the Submillimeter Array (SMA) and infer column densities for several species. In Section \ref{sec:obs}, we describe the observations, calibration, and data reduction. In Section \ref{sec:results}, we outline the methods of line detection and column density calculation, and present the results. In Section \ref{sec:disc}, we discuss these results and place them within the context of similarly massive, well-studied disks. In Section \ref{sec:concl}, we conclude with a summary of the study.

\section{Observations} 
\label{sec:obs}

\subsection{SMA Data Reduction \& Calibration}\label{sec:reduction}
We present SMA observations of GoHam from several separate spectral setups as outlined in Appendix \ref{sec:obs_appendix} in SMA projects 2020A-S033 (with IDs 9929 and 9969; PI: R. Teague) and 2020B-S057 (with ID 10613; PI: C. Qi).
All observations had SMA pointing center coordinates J2000 18h09m13.37s -32\degr10\arcmin49.5\arcsec with reference velocity offsets of –2.8 km\,s$^{-1}$.
One single track was collected for the SMA’s `Extended’ (EXT) configuration, and two tracks were collected for the SMA’s `Compact’ (COM) configuration, with the aim of maximizing the signal-to-noise on higher spatial frequencies, while minimizing sensitivity losses to larger angular scales. 
These SMA configurations provide projected uv-baselines that span 16.4--226.0\,m, over a total of $\sim641$ min (on-source). 
Total frequency coverage is approximately $213.5\,$GHz to $288.5\,$GHz, with small gaps at some frequencies (see Fig. \ref{fig:figure1}). 

Standard flux, bandpass and gain calibration sources were used, the latter referenced to the target location, each of which are detailed in Table~\ref{tab:calibrators}. 
To obtain products that could be calibrated in the \textit{Common Astronomy Software Applications} ({\tt CASA}) measurement set format \citep{2022PASP..134k4501C}, we converted each raw SMA data product with {\tt pyuvdata} version 3.1.3 \citep{Hazelton+2017, pyuvdata_Karto2025}, with a spectral re-binning factor of 2.
Auto-flagging and gain-phase, gain-amplitude, bandpass, and flux solution tables were calculated with the SMA `\textit{COMPASS}' (Calibrator Observations for Measuring the Performance of Array Sensitivity and Stability) tool, version 0.11.0 (Keating et al., in prep.). This routine follows that of \citet{Lovell2025b}, where the same calibration procedure was conducted to analyze the edge-on disk around IRAS~23077+6707. 
We manually checked for remaining interference spurs in all converted measurement sets with the {\tt CASA plotms} tool prior to applying the calibrator solutions to our target data.
We used the standard SMA reduction script\footnote{Which can be accessed via: \url{https://github.com/Smithsonian/sma-data-reduction}.} in {\tt CASA} v6.6.3, and used the task {\tt applycal} to apply the calibration tables to each measurement set.
We manually inspected the phases and amplitudes, as functions of both time and frequency separated by antenna, for all post-calibration target and calibrator data, which showed exceptional stability and band-to-band consistency.

With the {\tt CASA} task {\tt concat}, we combined the measurement sets, and split these into GoHam-only spectral regions for analysis purposes with the {\tt CASA} task {\tt mstransform}, which map to the lower- and upper-sidebands for both SMA receivers. No time-averaging nor spectral averaging was performed. No self-calibration was performed on the data.

\subsection{Line Imaging}
The continuum was removed from the data using the {\tt CASA} routine {\tt uvcontsub}. Continuum-subtracted data were imaged in {\tt CASA} using {\tt tCLEAN} until the noise level of the residuals was approximately equivalent to the noise of non-signal-containing regions (RMS on the order of 100 mJy beam$^{-1}$; see Table \ref{tab:molecules} for specific values). Cleaned data cubes were subsequently processed using the {\tt GoFish} package \citep{GoFish} to correct for the systemic velocity of $2.793 \pm 0.034$ km s$^{-1}$ \citep{2020MNRAS.495..451T} and to plot spectra for preliminary line identification. To maximize SNR, all identified line transitions were imaged at a downsampled channel width of 0.5 km\,s$^{-1}$. We specified a Briggs weighting scheme, a Hogbom deconvolver, and a \texttt{robust} parameter of 2.0. The resulting synthesized beam sizes, ranging from $\sim2\arcsec\times2\arcsec$ to $\sim4\arcsec\times3\arcsec$, are reported in Table \ref{tab:molecules}.

\subsection{Continuum Imaging}
The continuum data were processed in {\tt tCLEAN} with a Briggs weighting scheme, a Hogbom deconvolver, and a \texttt{robust} parameter of 0.5, and cleaned down to the theoretical sensitivity (on the order of 0.1 mJy). Using {\tt IMFIT}, a 2D-Gaussian fitting tool in {\tt CASA}, the integrated flux of the continuum was measured as $278\pm30$ mJy at 1.3 mm, 
consistent with previously-measured values \citep{2020MNRAS.495..451T, 2008A&A...483..839B}.
The uncertainty on this value is the quadrature sum of the statistical uncertainty produced by {\tt IMFIT} and the standard 10\% flux calibration uncertainty. This value is consistent within error bars across all datasets listed in Table \ref{tab:obs}. A map of the continuum is presented in the top panel of Fig. \ref{fig:figure2}.

\begin{table*}[ht]
    \caption{ Disk-integrated fluxes and additional parameters of the lines identified in GoHam.}
    \label{tab:molecules}
    \begin{threeparttable}
    \noindent\makebox[0.87\textwidth]{
    \begin{tabular}{@{}lccccccc@{}}
    \toprule
    \toprule
    Molecule  & $N'-N''$ & $J'-J''$           & $F'-F''$   & $\nu_0$ & Integrated Flux  & Beam & RMS \\ 
       &  &           &   & (GHz) & (Jy km s$^{-1}$) & ($\arcsec\times\arcsec$, $^{\circ}$)&  (mJy beam$^{-1}$) \\ \midrule
    H$_{2}$CO & -       & $3_{0,3}-2_{0,2}$ & -         & 218.222                                                 & 2.1$\pm$0.7                                                                & 1.90$\times$1.61, 3.0 & 60                                                     \\ \cmidrule(l){2-8}
     & -       & $3_{1,2}-2_{1,1}$ & -         & 225.698                                                 & 2.0$\pm$0.4                                                                & 4.48$\times$3.21, 156.9 & 90                                           \\ \cmidrule(l){2-8}
     & -       & $4_{1,4}-3_{1,3}$ & -         & 281.527                                                 & 2.5$\pm$1.9                                                                & 4.00$\times$2.19, 161.0 & 120                                           \\ \midrule
    C$^{18}$O & -       & $2-1$             & -         & 219.560                                                 & 6.6$\pm$1.1                                                               & 1.89$\times$1.60, 2.8 & 70                                            \\ \midrule
    $^{13}$CO & -       & $2-1$             & -         & 220.399                                                 & 19.2$\pm$2.0                                                               & 1.88$\times$1.59, 3.2 & 80                                          \\ \midrule
    C$^{17}$O & -       & $2-1$             & -         & 224.714                                                 & 3.3$\pm$0.4                                                                & 4.49$\times$3.22, 156.8  & 90                                           \\ \midrule
    CN        & $2-1$   & $3/2-1/2$         & $5/2-3/2$ & 226.660                                                 & 1.5$\pm$0.2\tnote{a}                                                                & 4.46$\times$3.20, 156.9  & 90                                          \\
              &         &                   & $1/2-1/2$ & 226.664                                                 & -                                                                & -                                            \\ \cmidrule(l){3-8} 
              &         & $5/2-3/2$         & $5/2-3/2$ & 226.874                                                 & 4.7$\pm$0.6\tnote{a}                                                              & 4.45$\times$3.19, 156.9   & 90                                         \\
              &         &                   & $7/2-5/2$ & 226.875                                                 & -                                                                & -                                            \\
              &         &                   & $3/2-1/2$ & 226.876                                                 & -                                                                & -                                            \\ \midrule
    $^{12}$CO & -       & $2-1$             & -         & 230.538                                                 & 47.5$\pm$4.9                                                               & 1.80$\times$1.52, 3.1 & 70                                           \\ \midrule
    CS        & -       & $5-4$             & -         & 244.936                                                 & 2.6$\pm$0.5                                                                & 4.29$\times$2.90, 157.5   & 90                                         \\ \midrule
    C$_2$H       & $3-2$   & $7/2-5/2$         & $4-3$     & 262.004                                                 & 2.0$\pm$0.4\tnote{a}                                                                & 4.29$\times$2.40, 161.0  & 110                                          \\
              &         &                   & $3-2$     & 262.006                                                 & -                                                                & -                                            \\ \cmidrule(l){3-8} 
              &         & $5/2-3/2$         & $3-2$     & 262.065                                                 & 0.8$\pm$0.3                                                                & 4.29$\times$2.40, 161.0   & 110                                         \\ \midrule
    HCN       & -       & $3-2$             & -         & 265.886                                                 & 5.1$\pm$0.6                                                               & 4.24$\times$2.37, 161.1       & 110                                      \\ \midrule
    HCO$^{+}$ & -       & $3-2$             & -         & 267.558                                                 & 8.0$\pm$0.9                                                                & 4.20$\times$2.35, 161.1         & 110                                   \\ \midrule
    DCO$^{+}$ & -       & $4-3$             & -         & 288.144                                                 & 0.2$\pm$0.3                                                                & 3.95$\times$2.14, 161.1  & 120                                                                                    
    \\ \bottomrule
    
    \end{tabular}
    }

    \begin{tablenotes}
    \item 
    \textbf{Notes:} The errors reported on the integrated fluxes are the quadrature sum of statistical uncertainties (the calculation of which is described in Section \ref{sec:fluxes}) and systematic uncertainty introduced by flux calibration, which is taken to be 10\%. Here $N$ is the rotational angular momentum quantum number excluding electron and nuclear spins, $J$ is the total angular momentum quantum number excluding nuclear spins, and $F$ is the total angular momentum quantum number. Molecules for which $F$ transitions are reported have observed hyperfine structure. All lines were imaged at a channel width of 0.5 km s$^{-1}$.
    \item[a] Blended lines.
    \end{tablenotes}
    \end{threeparttable}
\end{table*}

\subsection{Molecule Detections}
We detect multiple H$_2$CO, CN, and C$_2$H transitions (some hyperfine), four CO isotopologues, CS, HCN, HCO$^+$, and DCO$^+$. 
This chemical complexity is consistent with the population of molecules that have been detected thus far in disks at (sub)millimeter wavelengths  \citep[e.g.,][]{2023ARA&A..61..287O}. In total, 15 lines were identified in the data, including hyperfine structure. See Fig. \ref{fig:figure1} for a summary of the lines detected and the frequency coverage of this study.
Close-up spectra of all 15 detected lines, processed with {\tt GoFish}, can be found in Fig. \ref{fig:all_lines} of Appendix \ref{sec:all_lines}. Fig. \ref{fig:channel_maps} shows some individual channel maps of $^{12}$CO 2-1 emission. 

\section{Results} \label{sec:results}

\begin{figure*}[h]
    \centering
    \includegraphics[width=1.0\textwidth]{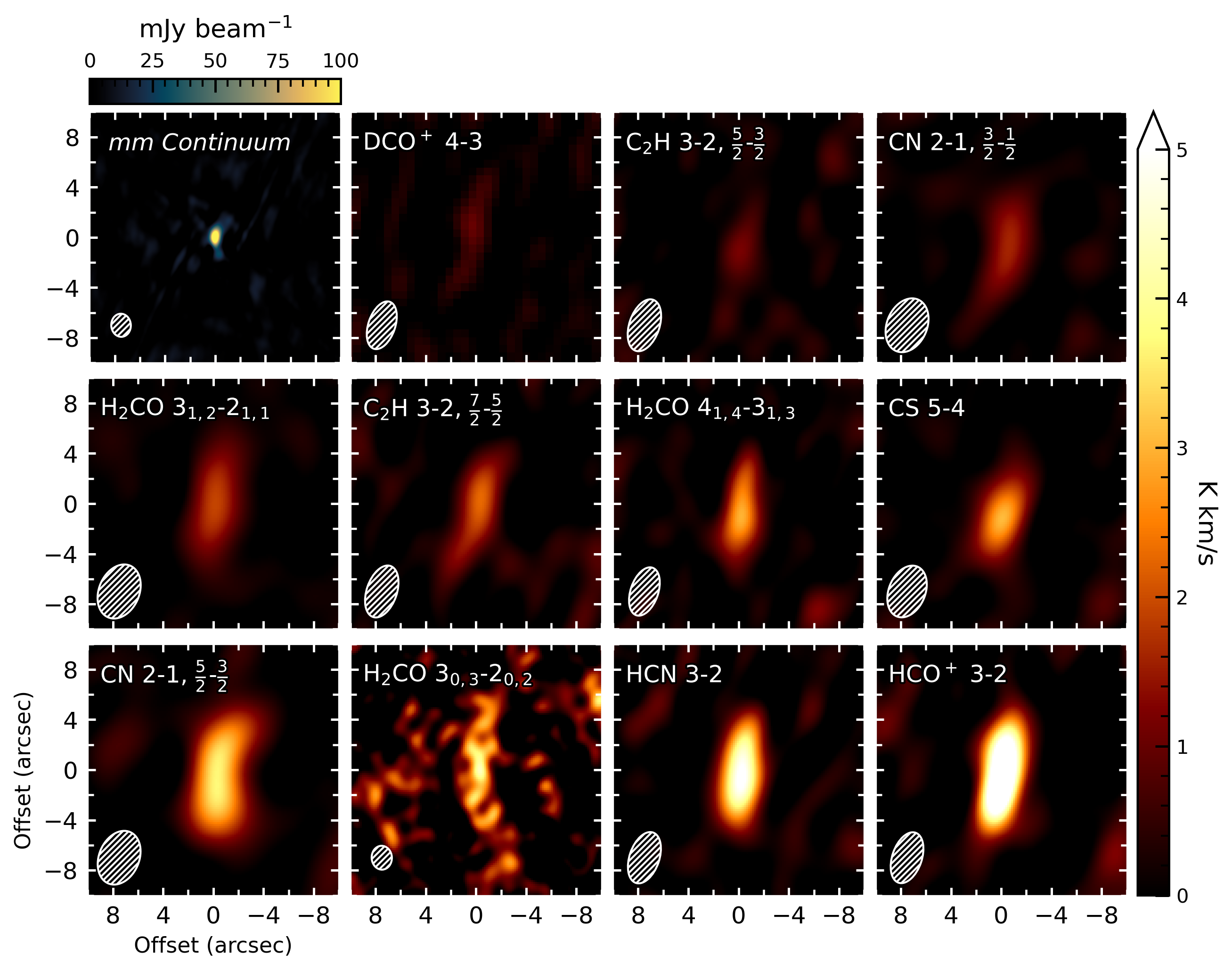}
    \includegraphics[width=1.0\textwidth]{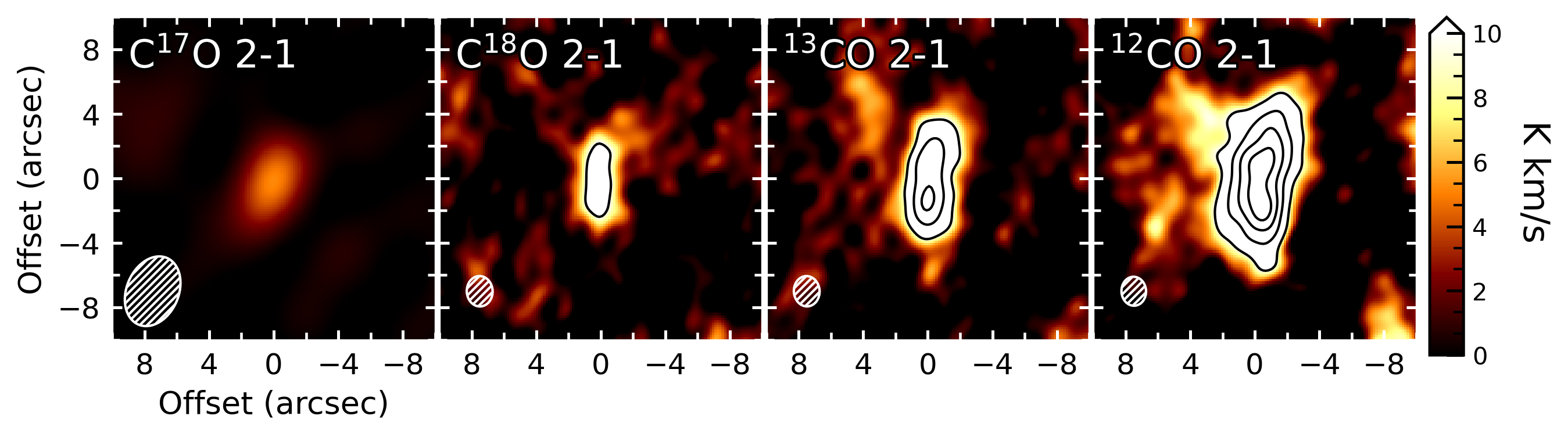}
    
    \caption{Continuum and zeroth moment maps showing the relative intensities of the lines detected in GoHam, ordered from faintest to brightest. The first panel of the top figure shows the 1.3 mm continuum emission, with a color bar cutoff at 100 mJy beam$^{-1}$ to highlight the continuum morphology. The dimmest lines are presented in the top figure. The bottom figure shows CO isotopologues with a color bar cutoff at 10 K km/s to emphasize the signal of the faintest lines. Contours are plotted starting at 10 K km/s and increasing in steps of 10 K km/s. Maps show brightness temperatures computed with the Rayleigh-Jeans approximation. The synthesized beam is shown in the bottom left corner of each panel.}
    \label{fig:figure2}
\end{figure*}

\subsection{Integrated Fluxes}\label{sec:fluxes}
For each of the lines reported in Table \ref{tab:molecules}, integrated fluxes were obtained by integrating over a broad elliptical mask. The mask was specified with a position angle of $175^{\circ}$ (the PA of the GoHam system), a major axis of $16\arcsec$, and a minor axis of $8\arcsec$ to best encompass the entire emission area, determined by a curve of growth analysis for $^{12}$CO 2-1 and $^{13}$CO 2-1. Both lines had clear maximum disk-integrated fluxes for this mask size. These lines were chosen as their spatial extents on a zeroth moment map are the greatest ($\sim12\arcsec$). The resulting conservative mask was applied to all lines to ensure no loss of true emission, but we note that this choice does introduce additional noise. We also note that the use of identical, large masks (assumed emission areas) for all lines means the column densities derived in our analysis are underestimates for molecules with a smaller extent than $^{12}$CO. Elliptical masks (rather than Keplerian masks) were used as GoHam's emission morphology is not well-represented by an inclined 2D disk; being viewed edge-on, GoHam's minor axis on the sky is dominated by its vertical extent. 

Statistical uncertainties on integrated fluxes were obtained for each line by offsetting the mask described above at spatial offsets of $-10\arcsec$, $0\arcsec$, and $10\arcsec$ along the horizontal axis and $-25\arcsec$, $0\arcsec$, and $25\arcsec$ along the vertical axis, as well as offsetting central velocity by 20 km\,s$^{-1}$ before, 10 km\,s$^{-1}$ before, and 10 km\,s$^{-1}$ after the range of channels containing emission. This resulted in a total of 27 samples of the noise in non-emission channels, integrated over an identical mask size and number of channels for each line. The standard deviation of each of these sets of flux measurements was taken to be the statistical uncertainty for each line. Systematic uncertainties from flux calibration are assumed to be $10\%$ of the measured fluxes, as is standard. The quadrature sum of these statistical and systematic uncertainties are given in Table \ref{tab:molecules} as the total uncertainties on the disk-integrated fluxes. In Fig. \ref{fig:figure2} we present a gallery of the integrated fluxes of each line, converted to K km\,s$^{-1}$ using the Rayleigh-Jeans approximation. 

We estimate the 50th-percentile recoverable flux scale ($\Theta_{\frac{1}{2}}$) of the data using Eq. A11 of \citet{1994ApJ...427..898W},
\begin{equation}
   \Theta_{\frac{1}{2}} = 8\farcs6 \bigg(\frac{\nu}{217 \text{ GHz}}\bigg)^{-1}\bigg(\frac{S_{\mathrm{min}}}{16\text{ m}}\bigg)^{-1},
\end{equation}
 where $\nu$ is frequency in GHz and $S_{\mathrm{min}}$ is the minimum projected baseline length in meters. Using a reference frequency of 230 GHz and $S_{\mathrm{min}}$ = 16.44 m, we find $\Theta_{\frac{1}{2}} \approx 7\farcs90$ for these observations. Based on single-channel maps (i.e. Fig. \ref{fig:channel_maps}), we find that the typical scale of disk emission is at most $\sim5\arcsec$-- smaller than the 50th-percentile recoverable flux scale of $\sim7\farcs90$. Thus, we do not expect there to be significant loss of flux to spatial filtering, or the addition of imaging artifacts. We note that, due to limited angular resolution, the shape of the emission is driven by the synthesized beam and thus is not particularly spatially resolved. For this reason, this study focuses on disk-integrated values. Possible limitations of this approach are discussed in Section \ref{sec:limitations}.

\subsection{Column Density Calculation} \label{sec: excitation}

We infer average column densities ($N$) for H$_{2}$CO, C$^{17}$O, CN, CS, C$_2$H, HCN, and HCO$^+$. We follow the methods first presented in \citet{1999ApJ...517..209G} and used in similar studies (e.g., \citet{2018ApJ...859..131L}, \citet{2018ApJ...864..133T}). We note than DCO$^+$ is not included in this calculation as the uncertainties on its integrated flux were too high to constrain its column density. We assume that all observed transitions are in local thermodynamic equilibrium (LTE) such that kinetic temperature is equal to excitation temperature ($T_{\mathrm{kin}} = T_{\mathrm{ex}}$). We caution that for radicals such as C$_2$H (and possibly CN), the emitting layer can be sufficiently tenuous that the lines are subthermally excited. In that case the inferred LTE $T_{\mathrm{ex}}$ should be interpreted as a rotational temperature and may underestimate $T_{\mathrm{kin}}$, biasing $N$. Previous disk studies have discussed this degeneracy and interpreted low C$_2$H excitation temperature as evidence for subthermal excitation and/or beam dilution \citep{2019ApJ...876...25B}. As discussed below, this study ultimately does not infer $T_{\mathrm{ex}}$, and instead adopts a range of excitation temperatures to infer $N$. Still, this nuance should be noted when interpreting our inferred column densities and their associated excitation temperatures. 

The Markov chain Monte Carlo (MCMC) random sampler {\tt emcee} \citep{2013PASP..125..306F} was used to model the posterior distribution of $N$ given known parameters of each transition (i.e. frequency, upper energy level, degeneracy, and Einstein-A coefficient for spontaneous emission) and our measured disk-integrated fluxes. Known values for the molecules were obtained from the JPL, CDMS, and LAMDA databases \citep{1998JQSRT..60..883P, 2016JMoSp.327...95E, 2005A&A...432..369S}. We include an optical depth correction following \citet{2018ApJ...859..131L}. We assume no line broadening beyond thermal, as the introduction of additional broadening was found to have a negligible impact on inferred column densities. 

For the molecules for which two or more lines are detected (i.e. H$_2$CO, CN, C$_2$H), we attempted to fit for $T_{\mathrm{ex}}$ alongside $N$, but we found that $T_{\mathrm{ex}}$ was unconstrained in all cases. This is likely due to the $N$-$T_{\mathrm{ex}}$ degeneracy (limited $E_u$ leverage and the fact that the disk-integrated fluxes sum emission over a broad range of radii and heights with different physical conditions), so for all species we report $N$ marginalized over an adopted $T_{\mathrm{ex}}$ range of excitation temperatures from 10 K to 50 K, following similar works \citep[e.g.,][]{2021ApJS..257....6G, 2025arXiv250926033D}. For CN and C$_2$H, one or more of the observed transitions contained blended hyperfine lines (see Table \ref{tab:molecules}), so molecule-specific models were made to account for this. These blended lines were modeled individually (unblended) and the sum of their modeled fluxes was compared to the observed blended fluxes. This follows standard practice \citep[e.g.,][]{2021ApJS..257...11B}.

\begin{table*}[ht!]
\centering
\caption{Column densities ($N$) and optical depths ($\tau$).}
\begin{threeparttable}
\begin{tabular}{ccccccc}
\toprule
\toprule
& &  & \multicolumn{2}{c}{T$_{\mathrm{ex}}=20$\,K} & \multicolumn{2}{c}{T$_{\mathrm{ex}}=50$\,K} \\
     Molecule & $J'-J''$ &$F'-F''$    & $\tau$ & $N$ (cm$^{-2}$ $\times 10^{12}$)& $\tau$ & $N$ (cm$^{-2}$ $\times 10^{12}$) \\ \toprule
H$_2$CO &$3_{0,3}-2_{0,2}$ & - & 0.31/0.21  &   $2.45^{+0.43}_{-0.43}$/$2.27^{+0.40}_{-0.41}$  &0.05/0.03 &$1.59^{+0.28}_{-0.29}$/$1.41^{+0.26}_{-0.26}$     \\ 
        &$3_{1,2}-2_{1,1}$ & - & 0.31/0.32   &     & 0.06/0.06&    \\  
        & $4_{1,4}-3_{1,3}$&  -  &0.32/0.33 &&0.09/0.09\\ \midrule

C$^{17}$O     &   $2-1$   & -  & 0.51& $3210^{+394}_{-398}$     & 0.11& $5020^{+611}_{-621}$      \\ \midrule

CN         & $3/2-1/2$ &  $5/2-3/2$  & 0.20& $6.14^{+0.58}_{-0.57}$   & 0.04 & $9.38^{+0.88}_{-0.90}$      \\
            &          & $1/2-1/2$           & 0.06 &&0.01\\  
           &$5/2-3/2$ & $5/2-3/2$ & 0.20&&0.04\\
           &          & $7/2-5/2$    &0.31&&0.07\\  
           &          & $3/2-1/2$    & 0.12&&0.03\\ \midrule
CS        & $5-4$      & - &0.34 & $2.78^{+0.56}_{-0.58}$   & 0.07 & $2.41^{+0.49}_{-0.50}$      \\ \midrule
C$_2$H    &  $7/2-5/2$    & $4-3$  & 0.13 & $8.45^{+1.55}_{-1.58}$  & 0.03 & $9.98^{+1.83}_{-1.83}$     \\
            &          & $3-2$     & 0.10 & &0.02\\  
        & $5/2-3/2$ & $3-2$& 0.10 & &0.02\\\midrule
HCN      &  $3-2$  & -  &  0.56 & $0.34^{+0.04}_{-0.04}$   & 0.12 & $0.44^{+0.05}_{-0.05}$      \\ \midrule
HCO$^+$    &   $3-2$  & - &   0.92  &$3.15^{+0.36}_{-0.37}$  &  0.19 & $3.70^{+0.42}_{-0.42}$      \\ \bottomrule
\end{tabular}
\begin{tablenotes}
    \item \textbf{Notes:} $\tau$ and $N$ values for H$_{2}$CO are presented for OPR$=2$/OPR$=3$.
\end{tablenotes}
\end{threeparttable}
\label{table:excit analysis}
\end{table*}

\begin{figure}[htb!]
    \includegraphics[width=0.47\textwidth]{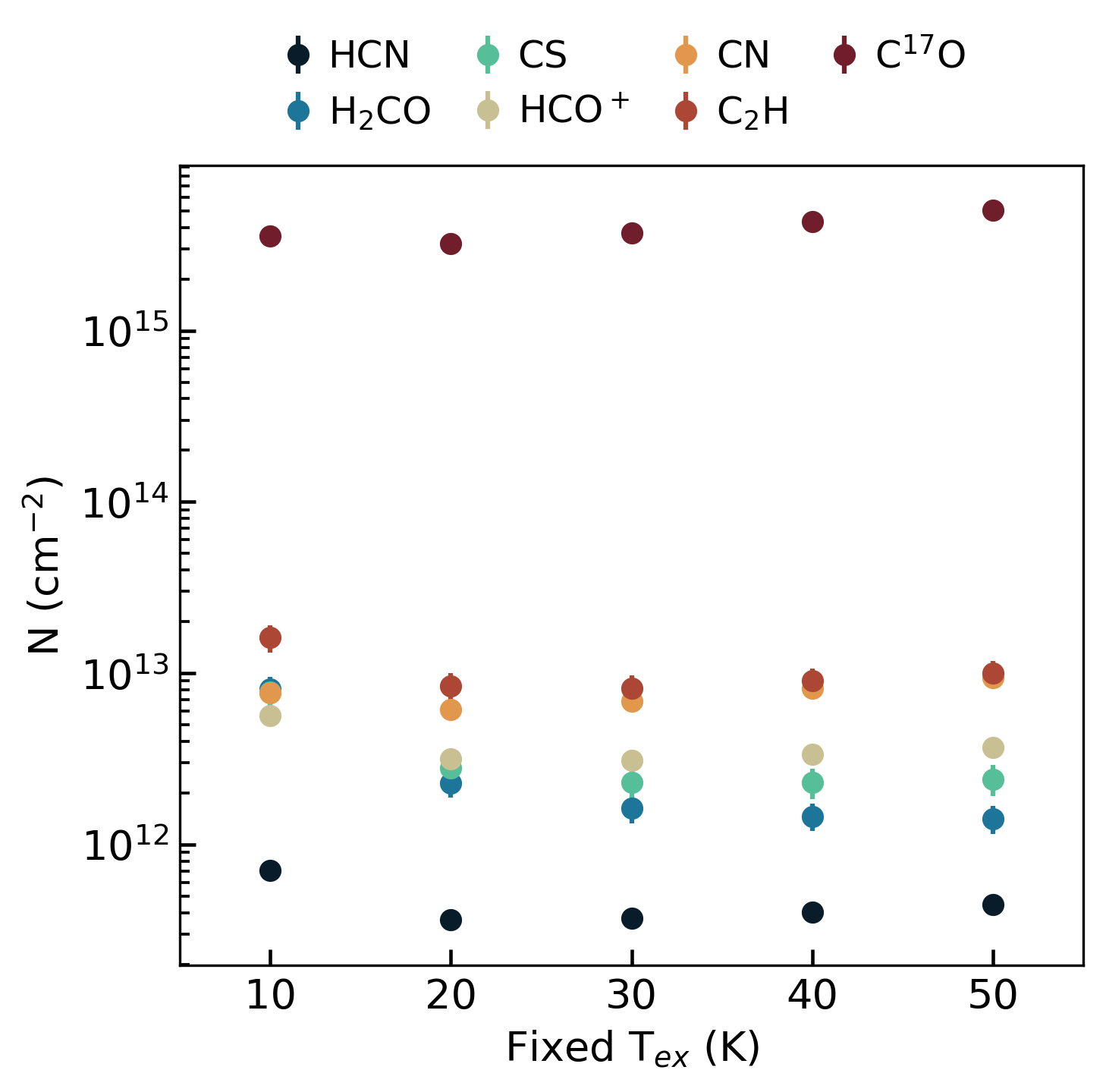}
    \caption{Disk-averaged column densities for HCN, H$_2$CO, CS, HCO$^+$, CN, C$_{2}$H, and C$^{17}$O, assuming fixed excitation temperatures of 10K, 20K, 30K, 40K, and 50K. For H$_{2}$CO, we present column densities calculated with a fixed OPR of 3.}
    \label{fig:figure3}
\end{figure}

We observed both ortho- and para-H$_{2}$CO transitions, and found that the ortho-to-para ratio (OPR) was also unconstrained in attempts to allow it to vary as a free parameter. In thermal equilibrium the OPR is expected to be $\sim3$ (the high-temperature limit), while the out-of-equilibrium OPR is expected to be $\sim2$ \citep{2011A&A...534A..49G}. Thus, we assume fixed OPRs of 2 and 3, following the precedent of similar works that have calculated or assumed an H$_{2}$CO OPR value of 2-3 to infer column densities \citep[e.g.,][]{1999ApJ...518..733D,2002A&A...381.1026R, 2021ApJ...906..111T, 2025arXiv250926033D}. We present column densities for both of these OPRs in Table \ref{table:excit analysis} and show that using an OPR of 2 rather than 3 has minimal impact on inferred column density. Fig. \ref{fig:figure3} shows that $N$ varies very little across the range of excitation temperatures used. 
The final $N$ values and uncertainties reported in Table \ref{table:excit analysis} are the median (50th) and 16th-to-84th percentile ranges of the posterior distributions, respectively.

\section{Discussion} \label{sec:disc}

\subsection{Chemical Complexity}
We detect a total of 15 individual lines in the data, including multiple transitions of H$_2$CO, CN, and C$_2$H (some hyperfine), four CO isotopologues (C$^{18}$O, $^{13}$CO, C$^{17}$O, and $^{12}$CO), CS, HCN, HCO$^+$, and a faint detection of DCO$^+$. This chemical complexity is consistent with the population of molecules that have been detected in disks at (sub)millimeter wavelengths, and we do not detect any unexpected species \citep{2023ARA&A..61..287O}. Furthermore, within the frequency ranges covered by the data, none of the molecules we expected to see were undetected. 
\subsection{Evidence for Changes Induced by Gravitational Instability}
The high mass and disk-to-star mass ratio of the GoHam system, as well as its Toomre $Q$ parameter of $Q\lesssim2$, may place it within the GI regime. Astrochemical models show that the dynamical processes within a gravitationally-unstable disk, i.e., shock heating, enhance the abundance of simple volatiles via desorption \citep{2011MNRAS.417.2950I}. These models, along with planetary heating models \citep{2015ApJ...807....2C}, suggest that H$_2$CO may be particularly sensitive to these processes (compared to other species such as CN). This is likely because H$_2$CO is formed in abundance via grain-surface processes, so more of it is available for desorption \citep{2015ApJ...809L..25L}. Thus, if GoHam is indeed gravitationally unstable, we may expect to see an elevated abundance of H$_2$CO compared to non-GI disks.

\begin{figure}[htb!]
    \includegraphics[width=0.47\textwidth]{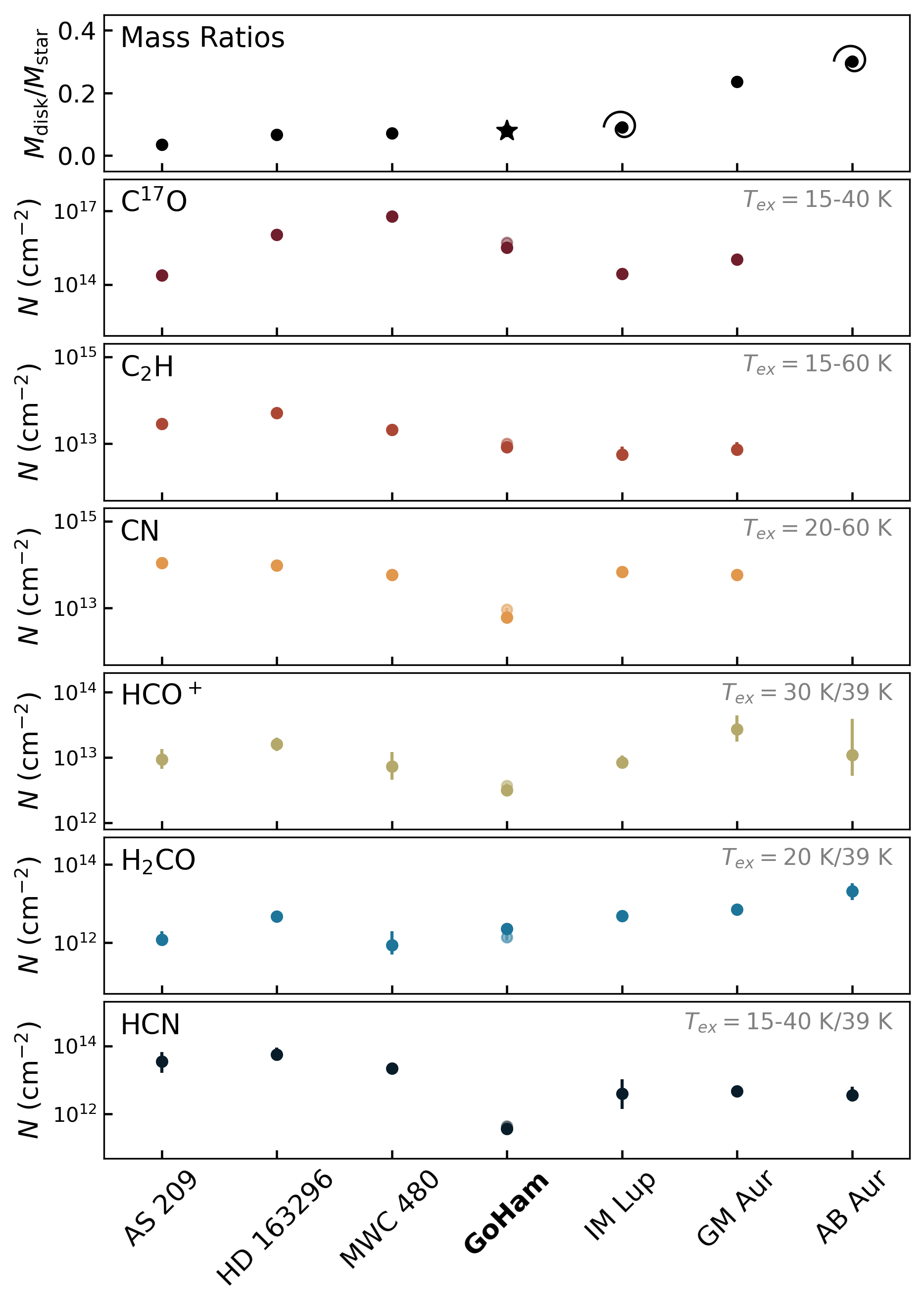}
    \caption{A comparison of disk-to-star mass ratios, and average column densities of GoHam with the MAPS disks and an additional known GI source. Disks are presented in order of increasing disk-to-star mass ratio. Spiral markers are used to indicate known GI sources and a star marker indicates GoHam in panel 1. The excitation temperatures associated with the given column densities are indicated in the top right of each panel; in some cases, {$T_{\mathrm{ex}}$} was allowed to vary as a free parameter, in which case we present the approximate range of temperatures obtained across all radii. When a fixed $T_{\mathrm{ex}}$ was used, we present the single value. In the cases for which data were available for AB Aur in addition to the MAPS disks, we present temperatures in the format $T_{\mathrm{ex, MAPS}}$/$T_{\mathrm{ex, AB Aur}}$. Note that $N$ values for CO, C$_2$H, and CN were not available for AB Aur. For GoHam, we present the column densities associated with $T_{\mathrm{ex}}=20$ K (opaque points) and $T_{\mathrm{ex}}=50$ K (transparent points). The H$_2$CO $N$ values for GoHam assume an OPR of 3.}
    \label{fig:column_density_comp}
\end{figure}

In Fig. \ref{fig:column_density_comp} we place GoHam in the context of several other well-studied disks -- the five disks in the Molecules with ALMA at Planet-forming Scales (MAPS) Large Program, and AB Aurigae (AB Aur). Of these sources, two are thought to be at least marginally gravitationally unstable: IM Lup \citep{2023MNRAS.518.4481L,2024NatAs...8.1148U} and AB Aur \citep{2024Natur.633...58S}. We compare the disk-to-star mass ratios of these seven sources \citep{ 2019ApJ...884...42S, 2020MNRAS.495..451T, 2021ApJS..257...18T,2023MNRAS.518.4481L, 2024AA...686A...9M, 2024Natur.633...58S}, as high disk-to-star mass ratios can be associated with a higher likelihood of GI. See Table \ref{tab:mass_ratio_values} for the current values we adopt. We also compare disk-averaged column densities for C$^{17}$O, C$_2$H, CN, HCO$^+$, H$_2$CO, and HCN when available \citep{2020A&A...642A..32R, 2021ApJS..257....6G, 2021ApJS..257...11B}. For C$^{17}$O, we compare column densities we obtained from our observations of GoHam to column densities for MAPS disks using the modeled column density profiles from \citet{2021ApJS..257....5Z}, which are consistent with full CO isotopologue ensemble. To obtain MAPS values for C$^{17}$O, we scaled  values by the local ISM ratio of $^{12}$CO/C$^{17}$O $=2005$ \citep{1999RPPh...62..143W}. We present column densities for GoHam inferred at excitation temperatures of $T_{\mathrm{ex}}=20$ K and $T_{\mathrm{ex}}=50$ K to demonstrate how the range of possible column densities compares to values obtained for these other disks.

\begin{figure}[htb!]
    \includegraphics[width=0.47\textwidth]{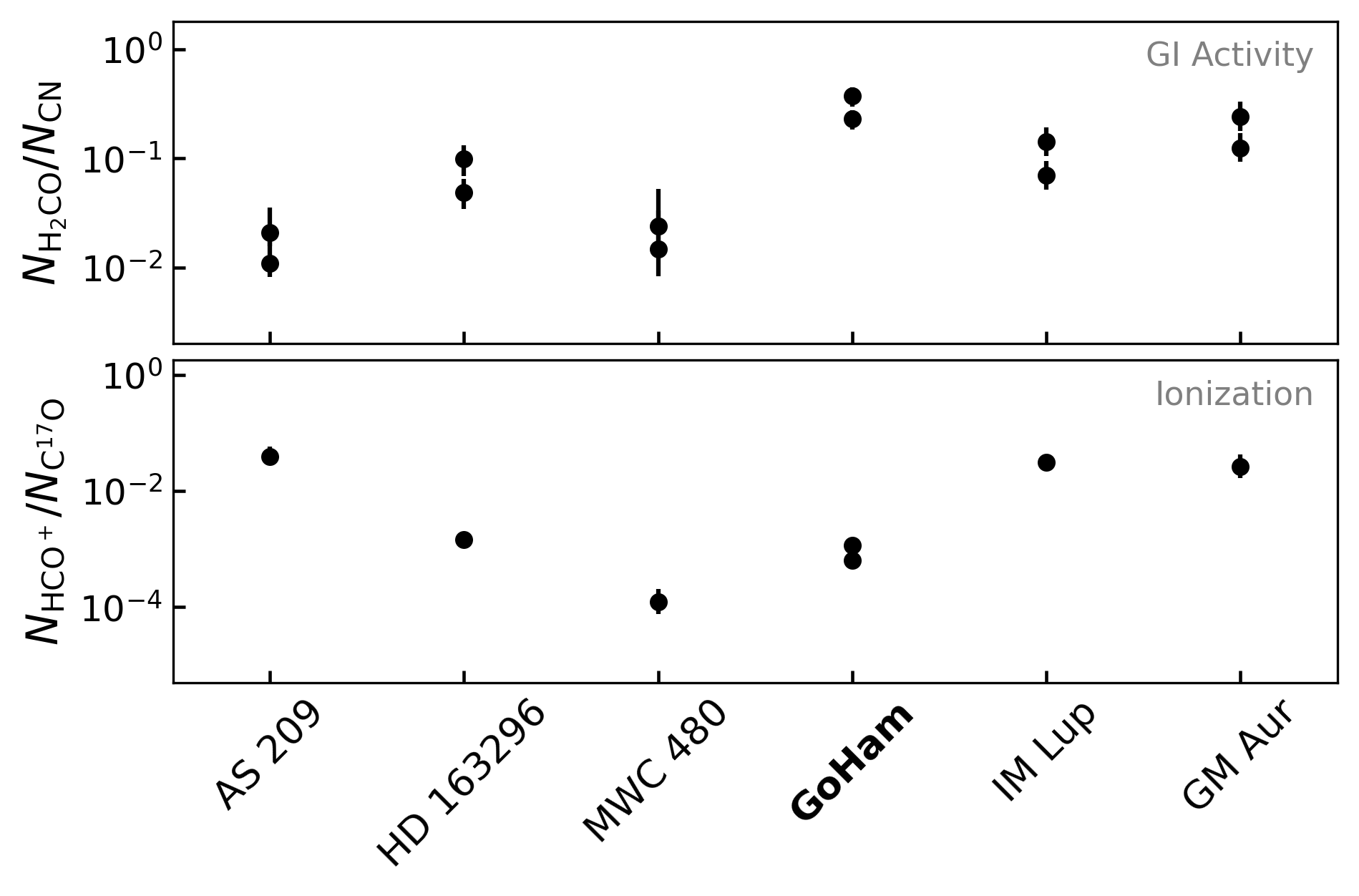}
    \caption{The top panel shows a comparison of disk-averaged H$_2$CO-to-CN column density ratios for GoHam and the MAPS disks. From the possible column densities available for different excitation temperatures, we present the highest and lowest ratios. These ratios are intended to probe H$_2$CO desorption resulting from GI activity in the disk. The bottom panel shows a comparison of disk-averaged HCO$^+$-to-C$^{17}$O column density ratios for the same disks. These ratios are intended to demonstrate comparative ionization rates within the disks. Disks are presented in order of increasing disk-to-star mass ratio.}
    \label{fig:column_density_ratios}
\end{figure}

As we will discuss in Section \ref{sec:limitations}, our use of disk-integrated values and integration over a large mask area for GoHam's highly inclined disk introduces some limitations. Thus, for a better one-to-one comparison with other disks, we also consider the ratio of disk-averaged column densities between H$_2$CO and CN (See Fig. \ref{fig:column_density_ratios}), as a means of probing the relative abundances of these species in each disk. (Here CN is used as an example; other molecules could be used to illustrate the same behavior.) From the possible column densities available for different excitation temperatures, we present the highest and lowest ratios. For MAPS disks, two sets of H$_2$CO column densities were available at $T_{\mathrm{ex}}=20$ K and $T_{\mathrm{ex}}=50$ K, while only one set of CN column densities were available. For GoHam, we used column densities for both species at $T_{\mathrm{ex}}=20$ K and $T_{\mathrm{ex}}=50$ K and present the highest and lowest ratios resulting from the four possible combinations of those values. It must be noted that H$_2$CO and CN typically trace different layers within the disk \citep[e.g][]{2010ApJ...722.1607W,2020A&A...636A..65G}, which our disk-averaging approach does not account for given our limited spatial resolution. Some variation in $N_\mathrm{H_2CO}$/$N_\mathrm{CN}$ values may be a result of these molecules tracing different layers within the disk. Higher-resolution studies that account for the differences in emitting layers between the molecules being compared may yield additional insights. Still, when comparing $N_\mathrm{H_2CO}$/$N_\mathrm{CN}$ values obtained for GoHam to values calculated for other disks using the same methods, we find that GoHam appears to have the highest ratio, ranging from $N_{\mathrm{H_2CO}}$/$N_{\mathrm{CN}}\approx1.5$ to $3.7$ depending on associated $T_{\mathrm{ex}}$. 
This ratio may indicate that H$_2$CO is enhanced in GoHam compared to other species, such as CN, that we expect to be less susceptible to shock heating. This enhancement, if truly present, may be an indicator that some dynamical processes associated with GI are present within GoHam \citep{2011MNRAS.417.2950I}. We note that this possibility would need to be verified with comparisons to abundance enhancements from shocks in a known GI-active disk whose disk-to-star mass ratio more closely resembles that of GoHam. We observe that GoHam has a higher disk-averaged $N_{\mathrm{H_2CO}}$/$N_{\mathrm{CN}}$ than IM Lup, a known gravitationally unstable disk. We suggest that this may be attributed to the fact that IM Lup does not have extensive spiral structure traced by the gas (which drives the shock-heating), so it may not be sufficiently unstable to experience shock heating to the degree that may be present in GoHam. This may result in reduced H$_2$CO desorption within the IM Lup disk.

As \citet{2021ApJS..257...13A} show, the column density ratio of HCO$^+$ to C$^{17}$O can be used to gain insight into the ionization rate within the disk. Although limited by our disk-averaged approach, we present $N_{\mathrm{HCO^+}}$/$N_{\mathrm{C^{17}O}}$ for GoHam and the MAPS disks in Fig. \ref{fig:column_density_ratios}, and show that the ratios for GoHam are most consistent with the lower-mass-ratio sources. This appears to indicate a standard ionization rate within GoHam, from which we cannot make any conclusions about the presence of GI. A radially-resolved study may provide insights into which regions of the GoHam disk experience comparatively higher ionization rates.

\subsection{Limitations of a Disk-Integrated Approach} \label{sec:limitations}
Given the limited angular resolution of the data, disk-integrated fluxes were used in calculating column densities. This method yields column densities for the entire disk and does not preserve information about the radial and vertical variations of the molecules. \citet{2021ApJS..257....9I} demonstrate the limitations of using disk-integrated values by performing both disk-integrated and radially resolved analyses to derive column densities and rotational temperatures for HC$_3$N,
CH$_3$CN, and $c$-C$_3$H$_2$ in four MAPS disks \citep{2021ApJS..257....1O}. They showed that their disk-integrated results were generally in agreement with previous observational results, but that for small radii ($\lesssim100$ au) their spatially-resolved analysis revealed significantly higher column densities for some large organic molecules. Similarly, it may be that the disk-averaged column densities shown in Table \ref{table:excit analysis} and Fig. \ref{fig:figure3} are underestimates at small radii, as species may become optically thick in the inner disk, but are good estimates at higher radii. 

Recent studies \citep[e.g.,][]{2020A&A...642L...7P} have noted vertical stratification between various molecules, such as $^{12}$CO and H$_2$CO. The large beam size of our observations introduces smearing effects that prevent the emitting layer from being well constrained. However, we qualitatively observe that $^{12}$CO emission in the disk is elevated compared to $^{13}$CO. Most other molecules we observe in the disk appear to be relatively consistent with the emission structure of $^{13}$CO, and more extended than C$^{18}$O, suggesting they trace a similar emission layer to $^{13}$CO (see Fig. \ref{fig:figure2}). Our analysis does not account for differences in vertical distribution of molecules. We note that this may introduce some additional uncertainty or bias into the resulting column densities we present. However, we do not expect this to significantly affect our conclusions given our disk-averaged approach.

We note that, if GI-induced enhancement of molecular abundances is sufficiently localized, disk-averaged values may not capture such enhancements; a followup study examining radial variation of H$_2$CO in GoHam may yield additional insights into the possibility of GI. We also caution that, given the large size of GoHam on the sky and our use of a large mask for calculating integrated fluxes, our $N$ values are averaged over a substantially larger area than the other disks we have discussed. This results in $N$ values that are generally low compared to these other disks. Finally, inferring disk-averaged values from disk-integrated fluxes over an edge-on disk such as GoHam may produce different results than the same process for a less-inclined disk, as the observed emission area is not as comparable to the true face-on area of the disk.

\section{Summary and Conclusions} \label{sec:concl}
We identify 15 molecular lines within GoHam's disk from a wideband survey using several SMA datasets taken over multiple years. We present the disk-integrated fluxes for these lines and infer column densities for several species. While the H$_2$CO-to-CN column density ratio may indicate the presence of gravitational instability affecting the gas-phase chemistry, our findings for GoHam are broadly consistent with other well-studied high-$M_{\mathrm{disk}}/M_{\mathrm{star}}$ disks. We find that no definitive conclusions can be drawn about GI based on the chemistry of GoHam without additional, higher-resolution data. 
Future studies with ALMA can further explore the chemical complexity and the possibility of GI in GoHam by targeting the molecules detected in this study at higher angular resolutions. These results establish an inventory of which molecules and transitions are present and bright within GoHam and establish order-of-magnitude estimates for the average column densities of several common molecules. This study lays the groundwork for future targeted studies that can offer insight into the conditions of planet formation in high-mass disks and the characterization of the gaseous over-density and giant protoplanet candidate GoHam b.

\vspace{5mm}
\section{Acknowledgments}
We thank the anonymous referee for feedback that greatly improved the quality of this study.
    
    The Submillimeter Array is a joint project between the Smithsonian Astrophysical Observatory and the Academia Sinica Institute of Astronomy and Astrophysics and is funded by the Smithsonian Institution and the Academia Sinica. We recognize that Maunakea is a culturally important site for the indigenous Hawaiian people; we are privileged to study the cosmos from its summit.

    EMC gratefully acknowledges MIT's Undergraduate Research Opportunity Program for funding supporting this study.

    TJH  acknowledges a Dorothy Hodgkin Fellowship, UKRI guaranteed funding for a Horizon Europe ERC consolidator grant (EP/Y024710/1) and and UKRI/STFC grant ST/X000931/1. Support for CJL. was provided by NASA through the NASA Hubble Fellowship grant No. HST-HF2-51535.001-A awarded by the Space Telescope Science Institute, which is operated by the Association of Universities for Research in Astronomy, Inc., for NASA, under contract NAS5-26555. 

    JDI acknowledges support from an STFC Ernest Rutherford Fellowship (ST/W004119/1).
    
    MRJ acknowledges funding provided by the Institute of Physics Belgrade, through the grant by the Ministry of Science, Technological Development, and Innovations of the Republic of Serbia.

\facility{SMA}

\vspace{5mm}

\bibliography{GoHamCitations}{}
\bibliographystyle{aasjournal}

\appendix
\section{Summary of Observations}
\label{sec:obs_appendix}

\begin{table}[h!]
\centering
\caption{Observation details.}

\begin{tabular}{@{}ccccccc@{}}    
\toprule \toprule
Project code & ObsID    & Configuration   &  Exp. time (min) & LO (RxA/RxB) & Obs. date (UT)\\ \midrule
2020A-S033   & 9929  & SMA EXT Track   & 205 min   & 341.0/225.5 &  2020 Jun 12, 07:00:20 - 15:51:55           \\ \cmidrule(l){2-6} 
             & 9969  & SMA COM Track   &     192 min       & 341.0/225.5  & 2020 Jul 02, 05:29:53 - 15:33:58             \\ \midrule
2020B-S057   & 10613 & SMA COM Track   &    244 min   & 272.5/234.0 &  2021 May 27, 07:44:39 - 16:55:54                 \\

             \bottomrule
\end{tabular}
\label{tab:obs}
\begin{tablenotes}
    \item \textbf{Notes:} All datasets have an intrinsic resolution of 139.648 kHz. Note that data centered at 340 GHz collected in some of these observations were not used in this study due to its lower quality. 
    \end{tablenotes}
\end{table}

\begin{table}[]
\centering
\caption{Calibrator details.}

\begin{tabular}{@{}ccccc@{}}
\toprule
\toprule
Project code & ID    &          & Calibrators &           \\ \midrule
             &       & Bandpass & Flux        & Gain      \\ \midrule
2020A-S033   & 9929  & 3c279    & Titan       & 1744-312  \\
             &       &          & Ceres       & 1924-292  \\
             &       &          & Neptune     &           \\ \cmidrule(l){2-5} 
             & 9969  & 3c454.3  & Titan       & J1744-312 \\
             &       &          & Ceres       & J1924-292 \\
             &       &          & Neptune     &           \\
             &       &          & Uranus      &           \\ \midrule
2020B-S057   & 10613 & 3c279    & Callisto    & 1744-312  \\
             &       & 3c454.3  &             & 1924-292  \\
             &       &          &             &           \\ \bottomrule
\end{tabular}
\label{tab:calibrators}
\end{table}

\newpage
\section{Lines Observed}\label{sec:all_lines}

\begin{figure*}[htpb!]
    \centering
    \includegraphics[width=\textwidth]{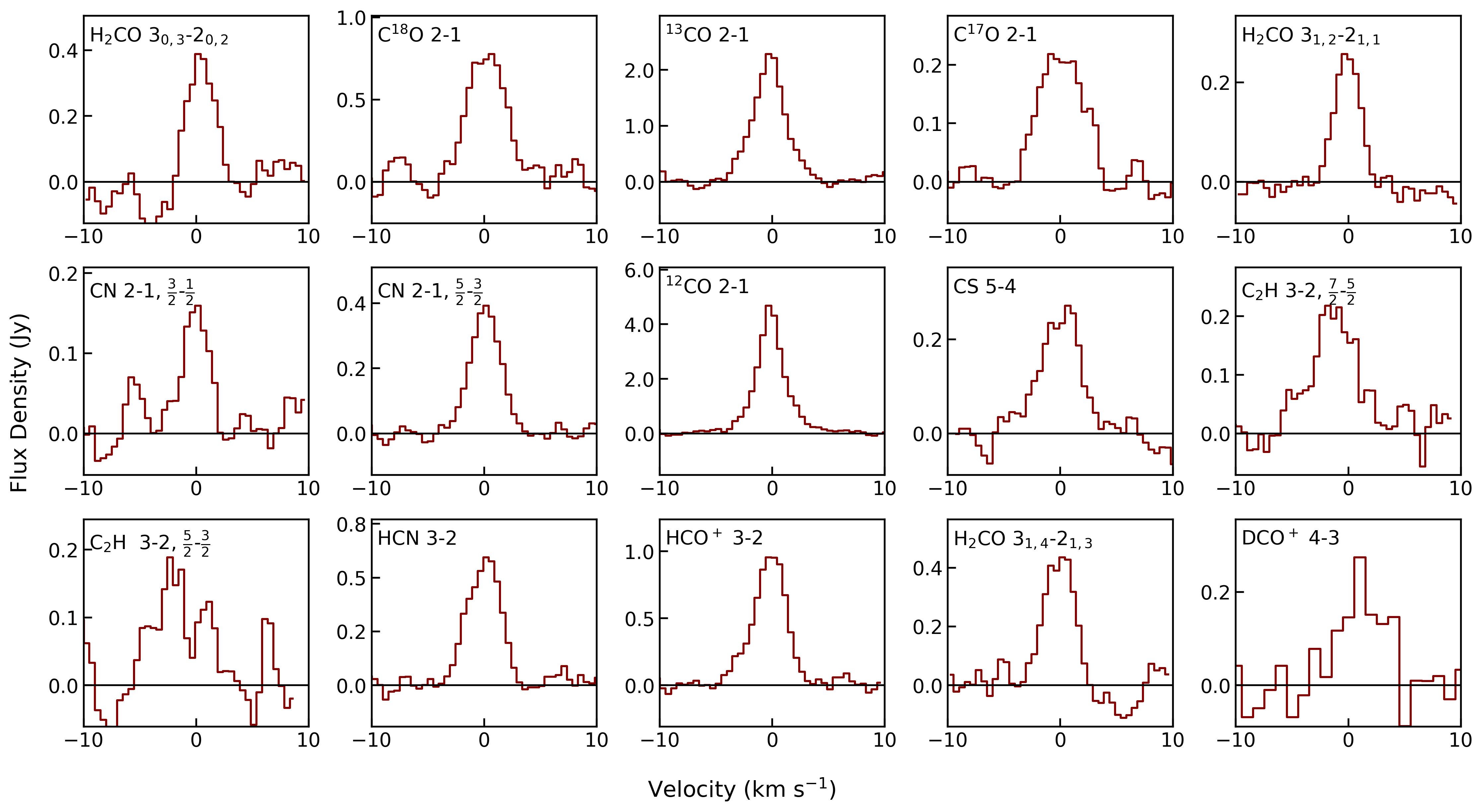}
    \caption{All individual molecular lines observed within the frequency ranges shown in Fig. \ref{fig:figure1}.}
   
    \label{fig:all_lines}
\end{figure*}

\begin{figure*}[htpb!]
    \centering
    \includegraphics[width=\textwidth]{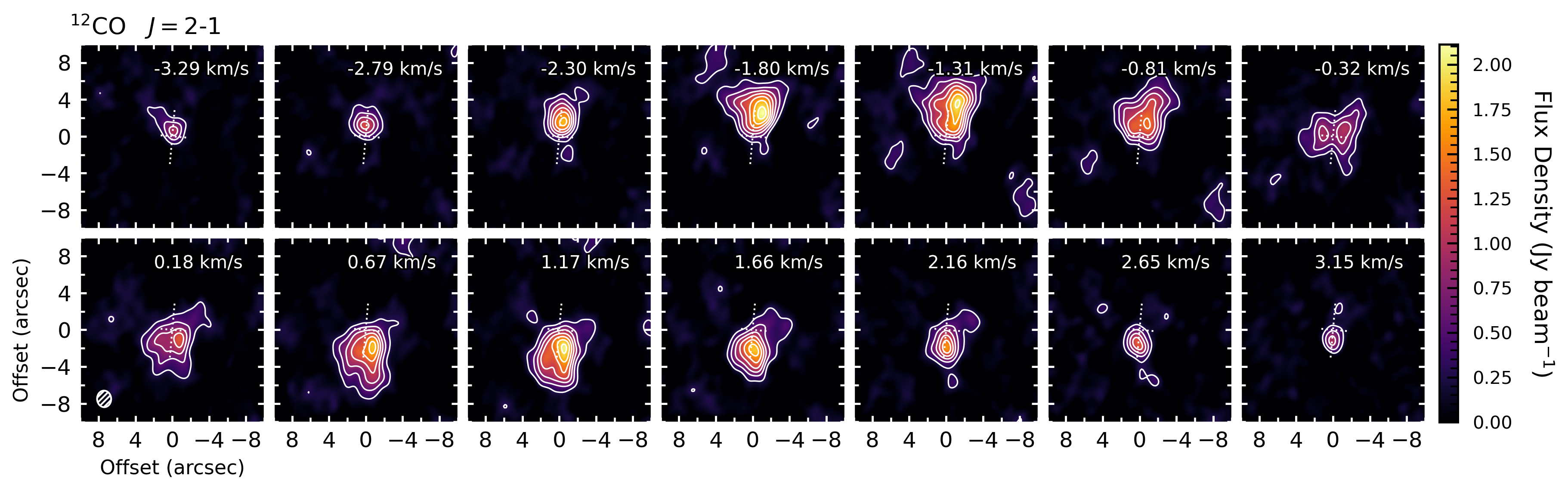}
    \caption{Channel maps of $^{12}$CO 2-1 emission. Contours are plotted starting at 4$\sigma$ and increasing in steps of 4$\sigma$, where $\sigma=70$ Jy beam$^{-1}$. The synthesized beam is plotted in the bottom left of the figure. Dotted lines show the orientation of GoHam's major and minor axes. The central velocity of each channel is shown in the top right of each panel.}
   
    \label{fig:channel_maps}
\end{figure*}
\newpage
\section{Disk-to-Star Mass Ratios}
\begin{table}[h]
\centering
\caption{Disk-to-star mass ratios used in the comparison discussion.}
\begin{tabular}{@{}ccc@{}}
\toprule
\toprule
Source    & $M_{\mathrm{disk}}/M_{\mathrm{star}}$ & Reference \\ \midrule
AS 209    & 0.04              & \citet{2021ApJS..257...18T, 2024AA...686A...9M}         \\
HD 163296 & 0.07              & \citet{2021ApJS..257...18T, 2024AA...686A...9M}         \\
MWC 480   & 0.07              & \citet{2019ApJ...884...42S, 2024AA...686A...9M}         \\
GoHam     & 0.08               & \citet{2020MNRAS.495..451T}                              \\
IM Lup    & 0.09              & \citet{2021ApJS..257...18T, 2023MNRAS.518.4481L}         \\
GM Aur    & 0.24             & \citet{2021ApJS..257...18T, 2023MNRAS.518.4481L}         \\
AB Aur    & 0.3                & \citet{2024Natur.633...58S}                              \\ \bottomrule
\end{tabular}
\label{tab:mass_ratio_values}
\end{table}

\end{document}